\documentclass[english,aps,pra,reprint,showkeys,noshowpacs,superscriptaddress]{revtex4-1}   
\usepackage[T1]{fontenc}	
\usepackage[latin9]{inputenc}	
\usepackage{geometry}		
\usepackage{graphicx}
\usepackage[above,below]{placeins}	
\usepackage{times}
\usepackage{hyperref}  
\hypersetup{colorlinks=true,urlcolor=blue,citecolor=blue,linkcolor=blue}   
\begin{document}

\title{Assessing a GTA professional development program}
\author{Emily Alicea-Mu\~{n}oz}
\affiliation{School of Physics, Georgia Institute of Technology, 837 State Street, Atlanta, GA, USA, 30332-0430}
\author{Joan Espar Masip}
\affiliation{Facultat de Matem\`{a}tiques i Estad\'{i}stica, Universitat Polit\`{e}cnica de Catalunya, C.~Pau Gargallo 14, Barcelona, Spain, 08028}
\author{Carol Subi\~{n}o Sullivan}
\affiliation{Center for Teaching and Learning, Georgia Institute of Technology, 266 4th Street NW, Atlanta, GA, USA, 30332-0383}
\author{Michael F. Schatz}
\affiliation{School of Physics, Georgia Institute of Technology, 837 State Street, Atlanta, GA, USA, 30332-0430}

\begin{abstract}
For the last four years, the School of Physics at Georgia Tech have been preparing new Graduate Teaching Assistants (GTAs) through a program that integrates pedagogy, physics content, and professional development strategies. Here we discuss various assessments we have used to evaluate the program, among them surveys, GTA self-reporting, and end-of-semester student evaluations. Our results indicate that GTAs who participate in the program find its practical activities useful, feel better prepared for teaching, make use of learner-centered teaching strategies, and receive higher scores in teaching evaluations.\end{abstract}


\maketitle

\section{Introduction}
Graduate teaching assistants (GTAs) are essential members of the teaching staff for large-enrollment introductory physics courses. Students in these courses spend nearly as much of their in-class contact time with GTAs (e.g., labs, recitations) as they do with faculty \cite{gardner}. Given their potentially large impact on student learning, and knowing that research shows that instructor preparation has a positive impact on teaching effectiveness \cite{gibbs,boman,holmes}, it is crucial to provide GTAs with adequate preparation and support for their roles as educators (see \cite{jossem} for several resources on GTA preparation in physics). Additionally, since teaching experience improves graduate students' research skills \cite{feldon}, a comprehensive preparation program for new GTAs can also act as an early step in their professional development as physicists. 

In 2013, the School of Physics at Georgia Tech began offering a one-semester GTA preparation course for first-year PhD students, developed based on the model described in \cite{carol}. To date, 92 graduate students have participated in the course. Preliminary review of course assessments indicated the program was well-received and was having a positive impact on the participants. Here we explore some of the assessments in detail to have a more rigorous understanding of the success and effectiveness of the program. Our results can provide the Physics Education Research (PER) community with information about effective methods for GTA professional development.

\section{Physics GTA Preparation}
\begin{figure}
\includegraphics[width=0.8\linewidth]{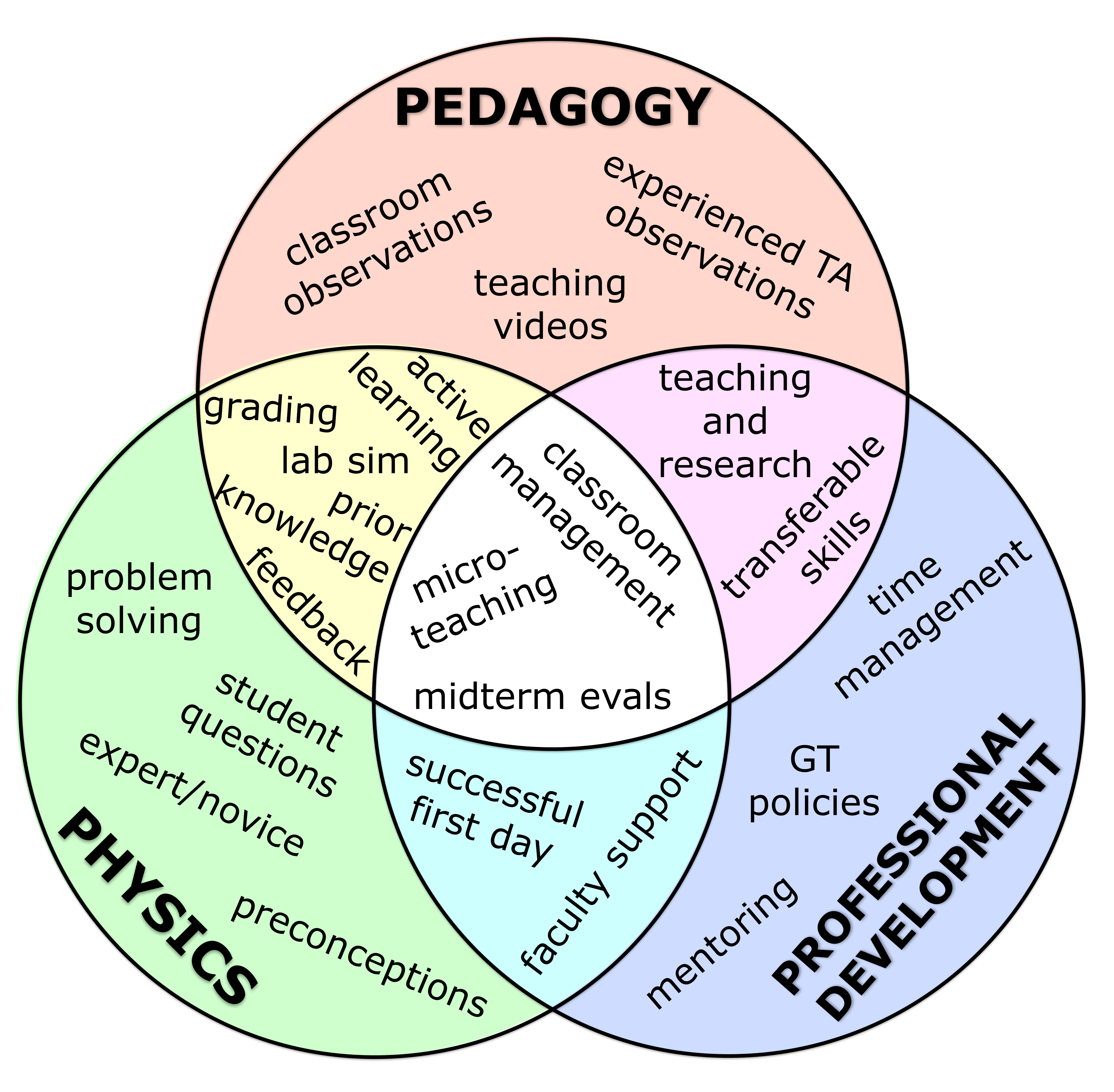}
\caption{Visualization of topics in our Physics GTA Preparation course (2016), emphasizing the integration of pedagogy, physics content, and professional development strategies. \label{venn}}
\end{figure}

The GTA preparation course has been taught every Fall semester since 2013. Over the years, the course contents have become more comprehensive and balanced, effectively integrating pedagogy, physics, and professional development (see Fig.~\ref{venn}). Although the details of the course contents have evolved, the main goals for the course remain the same: to help graduate students develop and apply learner-centered teaching practices, give and receive feedback on their teaching, manage classroom dynamics, and identify transferable skills they can use in their future careers as professional physicists. One of the researchers (EAM), a senior graduate student with significant GTA experience and involvement in PER and educational development, is the course instructor.

The course is structured in two parts: \textit{Orientation} and \textit{Follow-Ups}. The Orientation is a series of workshops, spread over several days, that comprise approximately 70\% of the contact hours for the course. It takes place two weeks before GTAs begin their teaching duties. The first workshop introduces grad students to their GTA responsibilities. The second workshop (Teaching Physics) covers the topics of differences between experts and novices, addressing preconceptions, and facilitating problem-solving. The third workshop is about classroom management and student motivation. The Orientation is rounded up with the Microteaching and Lab Simulation activities, in which the GTAs take turns to practice with their peers how to facilitate a problem-solving session and a lab experiment, respectively, and receive feedback from each other and the course instructor. The Follow-Ups are one-hour meetings that happen every $\sim$2-3 weeks during the semester; some of the topics covered in them include grading, midterm evaluations, and identifying transferable skills. All class meetings are interactive, with GTAs engaging in activities that reinforce research-based teaching methods, such as think-pair-share exercises, group discussions, and problem-solving sessions. Additionally, starting in 2014, the course instructor has conducted classroom observations of the GTAs at least once during the semester (twice in 2016). Video recordings of the observations are used, with the GTAs' permission, as examples for discussion in later iterations of the course.

\section{Methodology}
GTA and program assessment happens at various points before, during, and after the semester. The first is the Orientation Survey, which happens at the conclusion of the Orientation portion of the course. In this assessment, GTAs are presented with several five-point Likert scale questions to evaluate the course contents and determine their self-confidence for teaching. We administer pre/post assessments at the start and end of the semester to measure GTAs' pedagogical understanding and attitudes about teaching. Here we will discuss one of these assessments, the Approaches to Teaching Inventory (ATI). This research-validated instrument \cite{ATI} consists of 16 five-point Likert scale items that measure the extent to which GTAs approach teaching from a teacher-focused or student-focused perspective. During the last Follow-Up meeting in the semester, we give GTAs a Final Survey to evaluate the full contents of the GTA preparation course and its instructor. Finally, at the end of the semester, undergraduates are asked to complete Student Evaluations of Teaching for their lab and recitation GTAs. The GTAs receive the results of these evaluations after the semester is over.

\begin{table}[bp]
\caption{Assessment data analyzed in this paper.\label{thedata}}
\begin{ruledtabular}
\begin{tabular}{cc}
\textbf{Assessment} & \textbf{Data available (years)} \\ 
\hline
Orientation Surveys & 2013 - 2016 \\
ATI Pre/Post & 2014 - 2016 \\
Final Surveys & 2013 - 2016 \\
Student Evaluations & 2011-2012 (no GTA prep); 2013-2015 
\end{tabular}
\end{ruledtabular}
\end{table}

In this paper we analyze the data we have collected from the Orientation Surveys, the ATI pre/post, the Final Surveys, and the Student Evaluations. We employ these multiple streams of data to evaluate the effectiveness of the course along different dimensions (e.g., content, delivery, attitudes). The data are summarized in Table~\ref{thedata}. We use IBM SPSS Statistics 24 to perform the majority of our data analysis, and SciPy in a minority of cases. Our study has approval from the Georgia Tech Institutional Review Board and we have secured informed consent from study participants when required.

\section{Analysis and Results}

\subsection{Orientation Surveys}
In the first year of our GTA preparation course (2013), we wrote a short survey to receive feedback from the GTAs at the end of the Orientation. The survey had five free-response questions and one five-point Likert scale question, ``How prepared do you feel for your first day of teaching, on a scale of 1 to 5?'' None of the 21 responses reported a preparedness level of 1 or 2. The majority (52.4\%) reported a level of 4, and the rest were split between 3 (28.6\%) and 5 (19\%).

\begin{figure}
\includegraphics[width=0.99\linewidth]{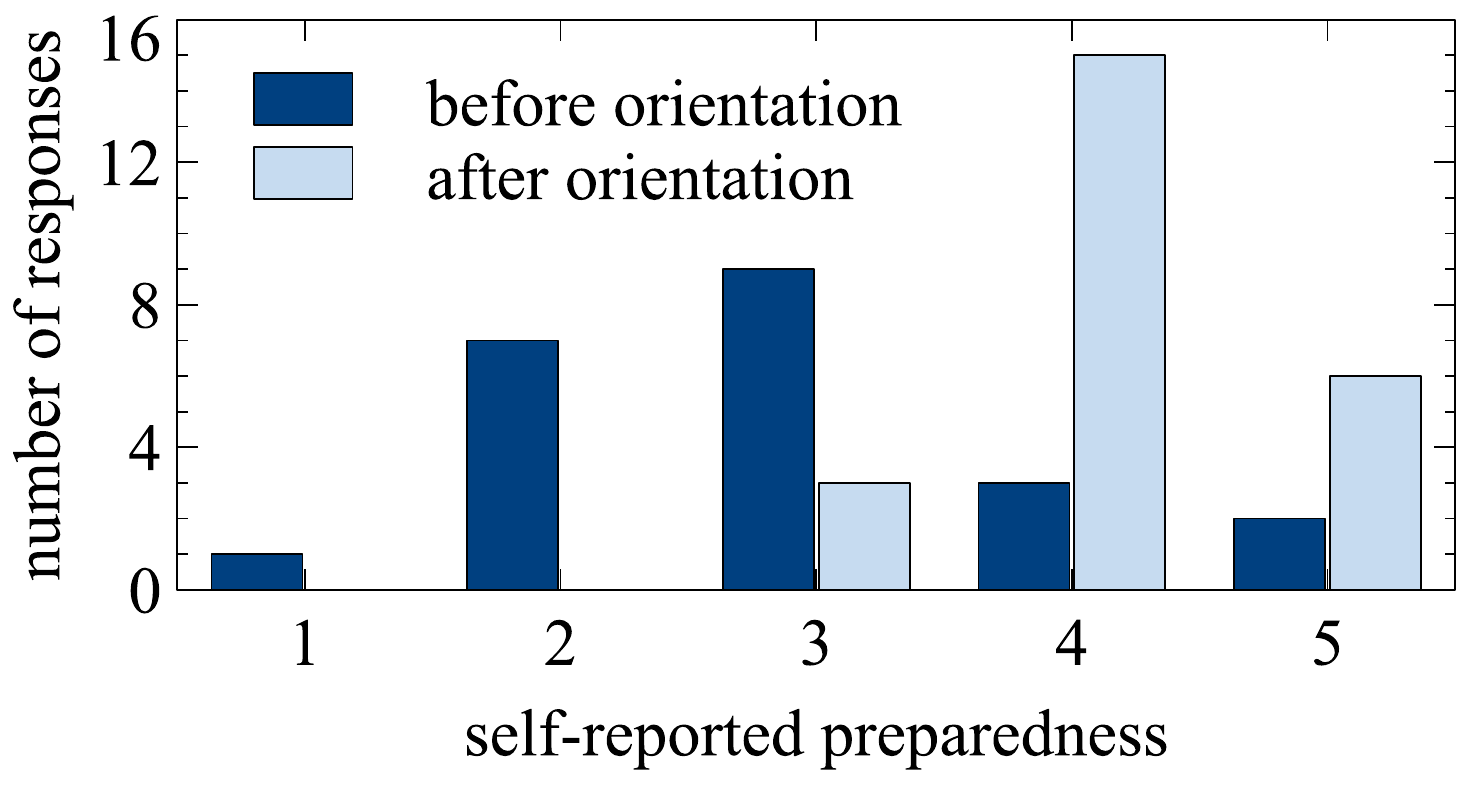}
\caption{Distributions of self-reported preparedness before and after the Orientation show that GTAs feel better prepared for teaching after going through the GTA preparation course. First-time GTAs in 2016 were asked to rate how prepared they felt for teaching before and after completing the Orientation portion of the GTA prep course, before the start of their teaching duties. The before ($N=22$) and after ($N=24$) distributions are noticeably different.\label{jumpstartplot}}
\end{figure}

In 2014 we rewrote the Orientation Survey to ask for more targeted feedback through five-point Likert scale statements. Although the statements varied from year to year, there were 10 common statements from 2014 to 2016. These statements explored GTAs' thoughts about the course content, its delivery, and their self-confidence about teaching. An analysis of variance revealed no statistical differences in the means of each separate statement across the three-year period. Each statement mean across three years demonstrates that GTAs in general find the Orientation to be useful (``Going through [Orientation] before the TA job begins is helpful to me,'' $M=4.54$, $SD=0.70$) and valuable (``Microteaching was a valuable practical experience,'' $M=4.54$, $SD=0.72$), and that they enjoy the interactive method in which the class is delivered (``I would have preferred more lecturing than activities,'' $M=2.13$, $SD=0.95$).

In 2016, GTAs were asked ``How prepared do you feel for your first GTA assignment at Georgia Tech?" (on a five-point Likert scale) before the Orientation started and again at the end of the Orientation. The responses allowed us to determine if GTAs' feelings of preparedness for teaching change after going through the first part of the GTA prep course. The results can be found in Fig.~\ref{jumpstartplot}. A non-parametric two-sample Kolmogorov-Smirnov (KS) test showed that GTAs feel significantly better prepared for teaching after participating in the Orientation ($D=0.65$, $p\ll0.001$)

\subsection{ATI Pre/Post}
\begin{figure}
\includegraphics[width=0.99\linewidth]{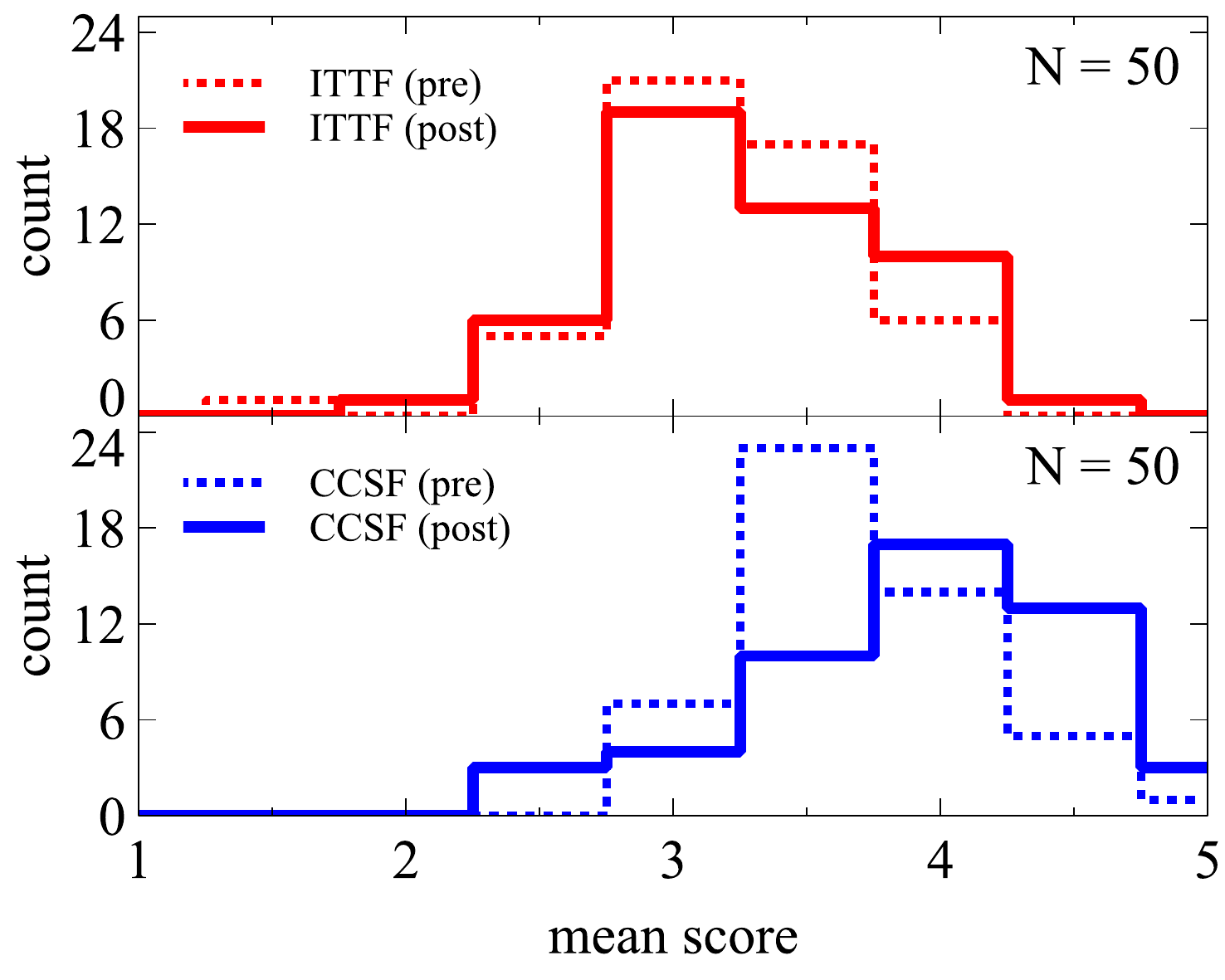}
\caption{GTAs' approaches to teaching are more student-focused after one semester of GTA preparation and teaching experience. Shown here are histograms of the ATI mean score distributions, at the start (pre, dotted lines) and end (post, solid lines) of the semester, for teacher-focused (ITTF, top, red) and student-focused (CCSF, bottom, blue) approaches to teaching. There is a pre/post statistical difference in the distributions and grand means of CCSF scores.\label{histograms}}
\end{figure}

We administer the ATI on the first and last class meeting. Of the 16 items in the ATI, eight explore Information Transmission/Teacher-Focused (ITTF) approaches, and the other eight explore Conceptual Change/Student-Focused (CCSF) approaches. For each GTA, we calculated the average of the eight ITTF items and the average of the eight CCSF items (teacher-focused mean and student-focused mean, respectively), in the pre-ATI and again in the post-ATI. The data is visualized in Fig.~\ref{histograms}. Using a non-parametric two-sample KS test we found no statistical difference in the distribution of mean scores for teacher-focused approaches; however, the distribution of student-focused mean scores is statistically different from pre to post ($D=0.28$, $p=0.032$). We then calculated the (pre and post) teacher-focused and student-focused grand means, by averaging all the teacher-focused means and all the student-focused means, respectively. Since the distributions of mean scores are not normal, we used a Wilcoxon signed-ranks test to determine whether there was a statistical difference in the pre/post grand means for teacher-focused and student-focused approaches. We found that the teacher-focused grand mean did not change significantly, but there was a significant change in the student-focused grand mean, from 3.52 to 3.72 ($Z=-2.278$, $p=0.023$).

\subsection{Final Surveys}
In 2013 and 2014, GTAs were asked at the end of the semester to list their top 3 most useful course topics. Microteaching topped the lists on both years. Midterm Evaluations, an activity in which GTAs both provide feedback for the course and receive feedback from their students, was also highly ranked, coming in at number 3 in 2013 and tying for first in 2014. The Grading Practice activity was among the top 3 in 2013, but absent from ranking the following year. The 2014 group selected Classroom Management as the second most useful topic, and Teaching Videos (in which GTAs discuss video clips of other GTAs' teaching) as third.

\begin{table}[bp]
\caption{Final Survey Top 3.\label{top3}}
\begin{ruledtabular}
\begin{tabular}{cc}
\textbf{Year} & \textbf{Top 3 (most useful)} \\ 
\hline
2013 & Microteaching, Grading, Midterm Evaluations \\
2014 & Microteaching/Midterm Evals, Class Mngmnt, Videos \\
2015 & Microteaching, Class Obs, Teaching Physics \\
2016 & Microteaching, Teaching Physics, Class Obs
\end{tabular}
\end{ruledtabular}
\end{table}

The 2015 and 2016 groups were given five-point Likert scale surveys to evaluate the usefulness of the course topics and activities. The 2015 group indicated Microteaching as being the most useful course activity ($M=4.38$, $SD=1.07$), followed by Individual Classroom Observations ($M=3.79$, $SD=1.29$) and Teaching Physics ($M=3.76$, $SD=1.06$). The 2016 group also considered Microteaching the most useful activity ($M=4.32$, $SD=0.72$), followed by Teaching Physics ($M=4.23$, $SD=0.69$) and Individual Classroom Observations ($M=4.09$, $SD=1.11$). A qualitative summary of these results can be found in Table~\ref{top3}.

\begin{figure*}
\includegraphics[width=0.99\linewidth]{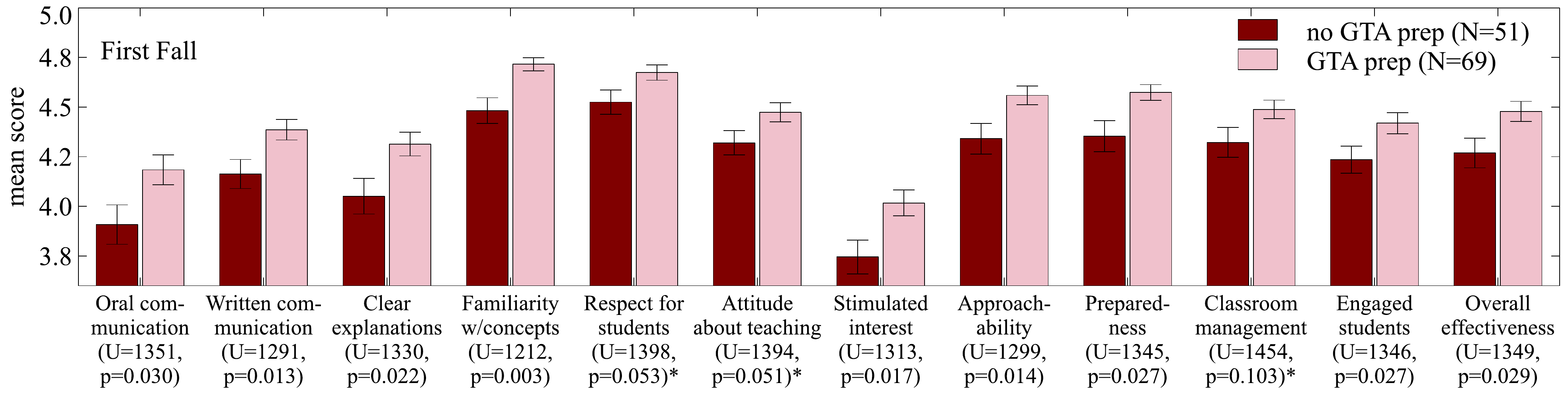}
\caption{GTAs who participated in the GTA preparation course receive higher student evaluation ratings at the end of their first semester of teaching than GTAs who did not participate in the preparation course. The differences are statistically significant in all categories except respect for students, attitude about teaching, and classroom management. Error bars indicate the standard errors of the means.\label{taosfig}}
\end{figure*}

\subsection{Student Evaluations}
The Georgia Tech Office of Academic Effectiveness conducts end-of-semester student evaluations of all instructors. The evaluation for GTAs consists of 12 five-point Likert items and three free-response questions. We gathered all available data for the Likert items, for the first fall semester of teaching, of all first-time physics GTAs between 2011 and 2015. The collected data was in the form of interpolated medians for each of the 12 Likert items, for each lab or recitation section taught by each GTA in that semester. From this we calculated each GTA's average for each of the 12 items; this way we had a single score per GTA for each Likert item. The distributions of scores for all  items were highly skewed towards the high end of the scale, which tells us that intro physics students at Georgia Tech are reluctant to give low ratings to their GTAs.

We consider participation in the GTA prep course as the intervention. GTAs whose first fall semester of teaching happened in 2011 or 2012 predated the GTA  course, so they are the no-intervention group ($N=51$), while the intervention group are those GTAs whose first fall semester of teaching happened in 2013 or later ($N=69$). Given the highly skewed nature of the distributions, we used a non-parametric Mann-Whitney test to determine if there were statistical differences between the two groups. We found that GTAs who participated in the GTA prep course receive higher student evaluation scores across the board. The differences are statistically significant for nine of the evaluation categories, and not statistically significant in three categories. These results can be found in Fig.~\ref{taosfig}, including all $U$ statistics and $p$-values. Non-significant results are indicated with an asterisk (*).

\section{Discussion and Conclusions}
Our GTA preparation course is well-liked and highly rated by the grad students who have participated in it. The course is effective at increasing GTAs' self-confidence in their teaching abilities. Informal conversations with a GTA Faculty Supervisor have reinforced this result, as we have been told they have seen far fewer ``freakouts'' from first-time GTAs at the start of the semester since the prep course went into effect.

The course has also been effective at increasing GTAs' learner-centered teaching approaches. In future work, we will explore the evolution of these approaches as the GTAs gain more teaching experience.

GTAs value the interactivity of the preparation course, and consider practical activities such as Microteaching, Midterm Evaluations, and Classroom Observations to be the most useful elements of the class. GTAs have also shown interest in learning more details about research-based teaching methods.

Students consistently give higher end-of-semester ratings to first-time GTAs who participated in the GTA preparation class than to first-time GTAs who had no formal teaching preparation. We can infer that participation in the GTA course is at least partially responsible for this difference, with participants being more effective first-time GTAs than non-participants. However, we must keep in mind the subjectivity of student evaluations, and all the arguments that have been made in favor and against them as valid assessments of teaching effectiveness and student learning \cite[e.g.,][]{moore,marsh,yunker,emery}.

The success of our GTA preparation course suggests that programs that integrate physics and pedagogy and emphasize the use of research-based teaching practices can be effective methods of GTA professional development.

\acknowledgments{We thank Ed Greco and Scott Douglas for valuable discussions on our research methods and results.}



\begin{thebibliography}{99}  

\bibitem{gardner} G.E. Gardner and M.G. Jones, Science Education, \textbf{20}, 31 (2011)
\bibitem{gibbs} G. Gibbs and M. Coffey, Active Learning in Higher Education, \textbf{5}, 87 (2004)
\bibitem{boman} J.S. Boman, Canadian Journal of Higher Education, \textbf{43}, 100 (2013)
\bibitem{holmes} N.G. Holmes, et al., The Physics Teacher, \textbf{51}, 218 (2013)
\bibitem{jossem} E.L. Jossem, American Journal of Physics, \textbf{68}, 502 (2000)
\bibitem{feldon} D.F. Feldon, et al., Science, \textbf{333}, 1037 (2011)
\bibitem{carol} T.T. Utschig, M.I. Carnasciali, C. Subi\~{n}o Sullivan, International Journal of Process Education, \textbf{6}, 3 (2014)
\bibitem{ATI} K. Trigwell and M. Prosser, Educational Psychology Review, \textbf{16}, 409 (2004)
\bibitem{moore} R. Moore, The American Biology Teacher, \textbf{52}, 260 (1990)
\bibitem{marsh} H.W. Marsh and L.A. Roche, American Psychologist, \textbf{52}, 1187 (1997)
\bibitem{yunker} P.J. Yunker and J.A. Yunker, Journal of Education for Business, \textbf{78}, 313 (2003)
\bibitem{emery} C.R. Emery, T.R. Kramer and R.G. Tian, Quality Assurance in Education, \textbf{11}, 37 (2003)

\end{thebibliography}

\end{document}